\documentclass[prl,aps,floats,letterpaper,floatfix,nofootinbib,superscriptaddress,preprintnumbers,twocolumn]{revtex4}
\usepackage{amsmath}
\usepackage{color}
\usepackage[mathenv]{}
\usepackage{graphicx}
\usepackage{natbib}
\begin{document}
\bibliographystyle{apsrev}

\newcommand{\oldtext}[1]{\textcolor{red}{\tt #1}}
\newcommand{\newtext}[1]{\textcolor{blue}{\sl #1}}

\title{Optical dilution and feedback cooling of a gram-scale
  oscillator to 6.9 mK}

\author{Thomas Corbitt}
\affiliation{LIGO Laboratory, Massachusetts Institute of Technology,
Cambridge, MA 02139, USA}
\author{Christopher Wipf}
\affiliation{LIGO Laboratory, Massachusetts Institute of Technology,
 Cambridge, MA 02139, USA}
\author{Timothy Bodiya}
\affiliation{LIGO Laboratory, Massachusetts Institute of Technology,
 Cambridge, MA 02139, USA}
\author{David Ottaway}
\affiliation{LIGO Laboratory, Massachusetts Institute of Technology,
Cambridge, MA 02139, USA}
\author{Daniel Sigg}
\affiliation{LIGO Hanford Observatory, Route 10, Mile marker 2,
Hanford, WA 99352, USA}
\author{Nicolas Smith}
\affiliation{LIGO Laboratory, Massachusetts Institute of Technology,
 Cambridge, MA 02139, USA}
\author{Stanley Whitcomb}
\affiliation{LIGO Laboratory, California Institute of Technology,
Pasadena, CA 91125, USA}
\author{Nergis Mavalvala}
\affiliation{LIGO Laboratory, Massachusetts Institute of Technology,
Cambridge, MA 02139, USA}

\begin{abstract}
  We report on use of a radiation pressure induced restoring force,
  the {\it optical spring effect}, to optically dilute the mechanical
  damping of a 1 gram suspended mirror, which is then cooled by active
  feedback (cold damping). Optical dilution relaxes the limit on cooling imposed by
  mechanical losses, allowing the oscillator mode to reach a minimum
  temperature of 6.9~mK, a factor of $\sim 40000$ below the
  environmental temperature. A further advantage of the optical
  spring effect is that it can increase the number of oscillations
  before decoherence by several orders of magnitude. In the present
  experiment we infer an increase in the dynamical lifetime of the
  state by a factor of $\sim 200$.
\end{abstract}

\definecolor{purple}{rgb}{0.6,0,1}
\date{\today}
\maketitle

To measure quantum effects in an oscillator, it is desirable to
prepare the system in a low energy state, such that the number of
quanta in the mode $N = E/\hbar \Omega_{\rm eff}$ is comparable to 1,
where $E$ is the energy of the mode and $\Omega_{\rm eff}$ is the
resonant frequency. Typically a macroscopic system is maintained far
above the quantum ground state by thermal fluctuations that enter
through its mechanical coupling to the environment, and drive its
motion.  An oscillator of mass $M$ and spring constant $K$ undergoes
motion at its resonant frequency $\Omega_{\rm eff} = \sqrt{K/M}$ that
is related to its effective (or {\it noise}) temperature $T_{\rm eff}$
by
\begin{equation}
  \label{eq:energy}
  \frac{1}{2} K x_{\rm rms}^2 = \frac{1}{2}k_B\, T_{\rm eff}\,.
\end{equation}
Reduction of the root-mean-squared motion $x_{\rm rms}$, and hence
$T_{\rm eff}$, may be achieved by a passive optical damping force
(``cavity cooling'')~\cite{Karrai, schwabNature2006,
  zeilingerNature2006, arcizetNature2006, schliesserPRL2006,
  harrisRSI2007}, or by an active feedback force (``cold
damping'')~\cite{manciniPRL1998, cohadonPRL1999,
klecknerNature2006,poggioPRL2007}. In either case, cooling is
possible because such a force imposes a non-mechanical coupling with
an external system that need not be in thermal equilibrium with the
environment.

The limit of these techniques occurs when the oscillator is critically
damped, placing an upper bound on the cooling factor at $Q_M$, the
mechanical quality factor of the oscillator.  However, in this Letter,
we show that the constraint on cooling is relaxed when radiation
pressure supplies the system's dominant restoring force, and
demonstrate experimentally a cooling factor that is larger than the
quality factor in the absence of radiation pressure.

In addition, it is desirable that a quantum state of the oscillator,
once prepared, should survive for more than one oscillation period,
enabling subsequent measurements to reveal quantum superpositions in
macroscopic objects~\cite{armour, bouwmeester}.  Interaction of the
quantum system with its noisy environment typically acts to produce
decoherence --- departure from an ideal coherent quantum
superposition. The thermal decoherence time of an oscillator subject
to mechanical viscous damping is given by~\cite{manciniPRL1998,
  bvtScience1980, cavesRMP1980, caldeiraleggettPRA1985}
\begin{equation}
  \label{eq:tau}
  \frac{1}{\tau} = \frac{\Gamma_M k_B T_M}{2\,\pi \hbar \Omega_{\rm eff}}\,,
\end{equation}
where $\Gamma_M$ is the mechanical damping constant of the oscillator,
and $T_M$ is the ambient temperature of the environment. In practice,
viscous mechanical damping and its associated thermal noise may not be
the only cause of decoherence.  For example, frequency and intensity
fluctuations of a laser beam used to measure the position of the
oscillator couple to its position, and could decohere the state.  To
include these effects generically, we extend Eq.~\ref{eq:tau}:
\begin{equation}
  \label{eq:extau}
  \frac{1}{\tau} = \frac{1}{2\,\pi \hbar \Omega_{\rm eff}}\displaystyle\sum_i \Gamma_i E_i = \frac{\Gamma_{\rm eff}k_B T_{\rm eff}}{2\,\pi \hbar \Omega_{\rm eff}}\,.
\end{equation}
The equation is written in terms of the characteristic energy,
$E_i$, and coupling, $\Gamma_i$, of each noise source to emphasize
that the noise need not be thermal in origin. The effective
temperature used here is the same as in Eq. \ref{eq:energy}. We
point out that in all cases $\Gamma_{\rm eff} T_{\rm eff} \geq
\Gamma_M T_M$, such that the energy flowing into the mode is never
decreased. So, in the best case, when no noise in addition to
thermal noise is present, the equality is satisfied; otherwise, the
inequality holds.

The average number of oscillations $n_{\rm osc}$ before decoherence is
\begin{equation}
  \label{eq:req} n_{\rm osc} = \frac{\hbar\Omega_{\rm eff}}{k_B T_{\rm eff}}
    \frac{\Omega_{\rm eff}}{\Gamma_{\rm eff}}\,.
\end{equation}
Unless $n_{\rm osc}$ exceeds unity, evidence of quantum superposition
is quickly buried under environmental noise. This ordinarily precludes
large objects from exhibiting such effects, since $\Omega_{\rm eff}$
tends to decrease for larger objects, due to their greater inertia. We
also note that non-mechanical damping techniques reduce $T_{\rm eff}$
of the mode by increasing $\Gamma_{\rm eff}$, while leaving
$\Omega_{\rm eff}$ and $n_{\rm osc}$ nearly unchanged.  Therefore the
mechanical oscillator must be fabricated so as to satisfy $n_{\rm osc}
> 1$ initially; this poses a significant experimental challenge that
increases with the size of the system.

Use of the optical spring effect in addition to non-mechanical damping
should allow a system to exhibit quantum behavior even though its
initial configuration does not satisfy $n_{\rm osc} > 1$.  The new
technique addresses two quantities of interest in measuring quantum
states of macroscopic objects: (i) the average motion of the object
($x_{\rm rms}$ in Eq.~\ref{eq:energy}), which is related to {\it state
  preparation}; and (ii) the number of oscillations of the mode before
the state decays ($n_{\rm osc}$ in Eq.~\ref{eq:req}), relating to {\it
  state survival}.

In order to reduce thermal motion, the mirror must be as weakly
coupled to the outside environment as possible, which in practice
requires that the mirror should be suspended, with the stiffness of
the suspension as soft as possible. A laser beam is used to create a
potential well by generating an optical restoring force, commonly
known as an optical spring~\cite{BCcontrols, 40mOS, corbittPIOS,
msupla2002os, Sheard}. This potential well creates a mode of
oscillation with a natural frequency of up to a few kilohertz. In
our experiment, the mode is dynamically unstable because of the
delayed optical response, but can be stabilized by application of
either electronic~\cite{corbittPIOS} or
optical~\cite{corbittPRL2007} feedback forces. The optical spring
shifts the oscillator's resonant frequency while leaving its
mechanical losses unchanged. The mechanical quality factor $Q_M$, as
limited by those losses, is increased by the factor $\Omega_{\rm
eff} / \Omega_M$, where $\Omega_M$ is the natural frequency of the
free mechanical oscillator. We refer to this as ``optical
dilution'', analogous to the phenomenon of ``damping dilution'' that
accounts for the fact that the $Q$ of the pendulum mode can be much
higher than the mechanical $Q$ of the material of which it is
made~\cite{saulsonPRD1990,dilution}. This mitigation of intrinsic
thermal noise is possible because a fraction of the energy is stored
in the (noiseless) gravitational field. In the case of the pendulum,
the dilution factor depends on the amount of elastic energy stored
in the flexing wire compared to the energy stored in the
gravitational field -- approximated by the ratio of the
gravitational spring constant to the mechanical spring constant. The
optical dilution introduced here accounts for the fact that thermal
noise in our mechanical oscillator is reduced due to energy stored
in the optical field (the optical spring force acts similar to the
gravitational force).

A further advantage of this scheme is that it does not conserve
$n_{\rm osc}$, because it changes the resonant frequency of the
oscillator by orders of magnitude. This should allow quantum effects
to become visible in a system that would not otherwise show them. We
note again that the coupling of thermal energy ($\Gamma_M T_M$) into
the oscillator is not decreased, but by raising the resonant frequency
$\Omega_{\rm eff}$, the amount of energy in a single quantum
increases, thereby increasing the decoherence time.

\begin{figure}[bth]
  \begin{center}
    \includegraphics[width=8cm]{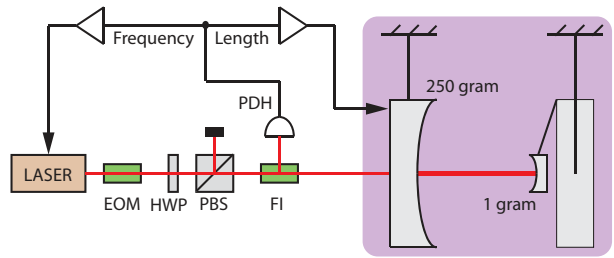}
    \caption{\label{fig:schematic} Simplified schematic of the
      experiment. About $100$~mW of $\lambda_0=1064$~nm Nd:YAG laser
      light passes through a Faraday isolator (FI) and a
      half-waveplate (HWP) and polarizing beamsplitter (PBS)
      combination that allows control of the laser power, before being
      injected into the cavity, which is mounted on a seismic
      isolation platform in a vacuum chamber (denoted by the shaded
      box).  A Pound-Drever-Hall (PDH) error signal derived from the
      light reflected from the cavity is used to lock it, with
      feedback to both the cavity length (actuated via magnets affixed
      to each suspended mirror), as well as the laser frequency.}
  \end{center}
\end{figure}

The experiment shown schematically in Fig.~\ref{fig:schematic} was
performed to demonstrate the optical dilution technique. The input
mirror of the $L=0.1$~m long cavity has mass of $0.25$~kg and is
suspended as a pendulum with oscillation frequency of $1$~Hz for the
longitudinal mode. The $10^{-3}$~kg end mirror is suspended by two
optical fibers $300$~$\mu {\rm m}$ in diameter, that are attached to
a stainless steel ring. The stainless steel ring is, in turn,
suspended as a 1~Hz pendulum. The oscillation frequency of the
longitudinal mode of the end mirror is $\Omega_M = 2\,\pi \times
12.7$~Hz, with quality factor $Q_M = 19950$, determined by measuring
the ringdown time of the mode. The input mirror transmissivity is
$\mathcal{T}_i = 800\times10^{-6}$, while that of the end mirror is
$10^{-5}$, and the laser wavelength is $\lambda_0 = 1.064 \times
10^{-6} {\rm~m}$. On resonance, the intracavity power is enhanced
relative to the incoming power by a resonant gain factor
$4/\mathcal{T}_i \approx 5 \times 10^3$, and with resonant linewidth
(HWHM) of $\gamma = \frac{\mathcal{T}_i\,c}{4\,L} \approx 2\,\pi
\times 95$ kHz.

At zero detuning from resonance, the stored power in the cavity exerts
a constant (dc) radiation pressure on each mirror. When the cavity is
detuned, changes in its radiation pressure give rise to both a
position-dependent restoring and a velocity-dependent damping force.
For a cavity with detuning $\delta$ and input power $I_0$, and change
of its length $x$, the radiation pressure force written in the
frequency domain is
\begin{equation}
  F = -Kx + M \Gamma \times\left(i\Omega x\right)\,,
\end{equation}
where the spring constant $K$ and damping coefficient
$\Gamma$ at each frequency $\Omega$ are given
by~\cite{corbittPRL2007}:
\begin{eqnarray}
  \label{eq:K} K\left(\Omega\right) &=& K_0\,\frac{\left[1 +
      \left(\delta/\gamma\right)^2 - \left(\Omega/\gamma \right)^2 \right]
  }{\left[1 +
      \left(\delta/\gamma\right)^2-\left(\Omega/\gamma\right)^2 \right]^2
    +
    4\,\left(\Omega/\gamma\right)^2 }\\
  \label{eq:damp} \Gamma\left(\Omega\right) &=& \frac{2\,K_0/\left(M
      \, \gamma \right)}{\left[1 +
      \left(\delta/\gamma\right)^2-\left(\Omega/\gamma\right)^2 \right]^2
    + 4\,\left(\Omega/\gamma\right)^2 }\,.
\end{eqnarray}
Here
\begin{eqnarray}
  \label{eq:K0} K_0 &=& \frac{2}{c}\frac{dP}{dL}=\frac{128\,\pi \,
    I_0\, \left(\delta/\gamma \right) }{\mathcal{T}_i^2 \,c \, \lambda_0
  }\left[\frac{1}{1+\left(\delta/\gamma\right)^2} \right] \,
\end{eqnarray}
and $M$ is the reduced mass of the two mirrors. The natural resonant
frequency of the system is shifted to
\begin{equation}
\Omega_{\rm eff} = \sqrt{\Omega_M^2 + K\left(\Omega_{\rm eff}\right)
/ M}.\label{eq:omegaOS}
\end{equation}

\begin{figure}[t]
  \begin{center}
    \includegraphics[width=9.0cm]{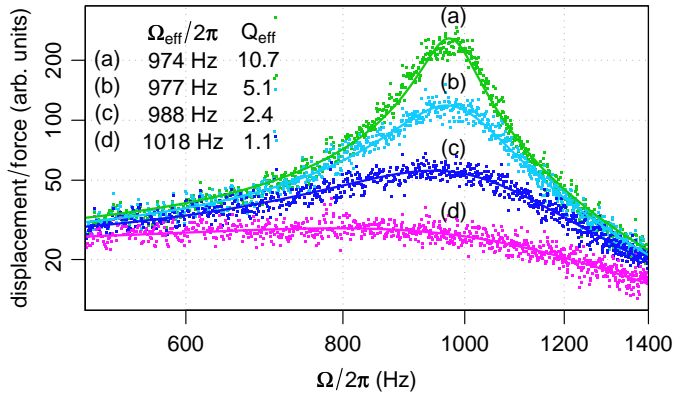}
    \caption{\label{fig:tf} The transfer function of an applied force
      to mirror motion, for increasing levels of damping
      [curves (a) to (d)]. The force is applied via the magnet/coil actuators,
      and the response is measured by the PDH error signal. The points are measured data,
      and the lines are fitted Lorentzians from which the resonant
      frequency and damping constant are derived for each
      configuration. Statistical errors in the fit parameters are of
      order 1\%.}
  \end{center}
\end{figure}

The cavity is locked off-resonance by $\delta \approx 0.5 \gamma$,
to maximize the optical restoring force. The error signal for the
locking servo, generated using the Pound-Drever-Hall
technique~\cite{PDH}, is split between a high bandwidth analog path
fed back to the laser frequency, and a digital path fed back to the
input mirror's magnet/coil actuators.  The digital feedback is used
at frequencies below $10$~Hz to keep the cavity locked in its
operating state. The analog feedback to the laser frequency is
arranged so that it damps and cools the motion of the oscillator, a
cold damping technique.  The effective damping may be controlled by
adjusting the gain of the feedback loop.  Additional analog feedback
is supplied to the magnet/coil actuators to damp a parametric
instability of the input mirror at
28~kHz~\cite{kippenbergPRL2005,corbittPIOS}.

By comparison with our previous report~\cite{corbittPRL2007}, in the
present experiment: (i) the cavity is shortened by a factor of 10 to
reduce the effect of laser frequency noise; (ii) the end mirror
suspension is reduced in stiffness by a factor of 180, and in
mechanical loss by a factor of 80; (iii) active feedback to the
laser frequency supplies the cold damping force, instead of a second
detuned optical field; and (iv) only 100~mW of incoming laser power
is used, which was necessary to avoid exciting a parametric
instability of the 1 gram mirror at 137~kHz.

\begin{figure}[t]
  \begin{center}
    \includegraphics[width=9.0cm]{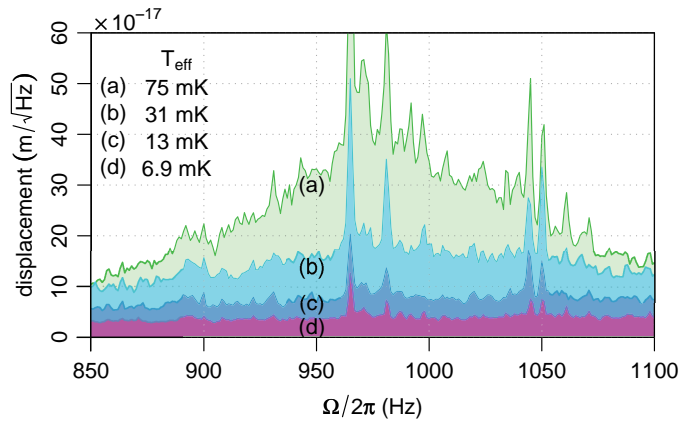}
    \caption{\label{fig:cool} The measured noise spectral density of
      the mirror displacement. The curves (a) to (d) correspond to
      increasing gain in the damping feedback loop; for each, the
      parameters of the resonance are measured and depicted in
      Fig.~\ref{fig:tf}. The spectra are integrated from 850 to 1100
      Hz, the frequency range where the mirror motion is the dominant
      signal, to obtain the rms motion of the mirror and its effective
      temperature. The broad limiting noise source is frequency noise
      of the laser. Narrow spectral features in addition to the main
      optical spring resonance are due to coupling of acoustically
      driven phase noise.}
  \end{center}
\end{figure}

\noindent {\bf Cooling:} The noise in our experiment remains dominated
by frequency noise of the laser at $\Omega_{\rm eff}$. We estimate the
effective temperature of the optomechanical mode, as determined by
this noise, according to Eq.~(\ref{eq:energy}).

To determine $x_{\rm rms}$ in our experiment, we first find the
resonant frequency and damping of the oscillator by measuring its
frequency dependent response to a driving force, shown in
Fig.~\ref{fig:tf}. In the same configuration, we then measure the
noise spectral density of the error signal from the cavity, calibrated
by injecting a frequency modulation of known amplitude at 12~kHz.  The
measured displacement spectra, as the electronic damping was varied,
are shown in Fig.~\ref{fig:cool}. Since the optical spring resonance,
given by Eq.~(\ref{eq:omegaOS}), is at $\Omega\approx 2\,\pi \times
1000$ Hz, we integrate the spectrum from $850$~Hz to $1100$~Hz to
obtain an estimate of the motion of the mirror. At other frequencies,
sensing noise not present on the mirror itself is dominant. To correct
for the finite integration band, we assume a thermally driven
displacement noise spectrum for the oscillator, given by
\begin{eqnarray}
  \label{eq:xrms} \left<x^2\right> = \frac{4\,k_B\,T_{\rm eff}\,\Gamma_{\rm
      eff}/M}{(\Omega_{\rm eff}^2 - \Omega^2)^2 + (\Omega_{\rm
      eff}\,\Omega/Q_{\rm eff})^2}\,,
\end{eqnarray}
and find $T_{\rm eff}$ by setting our measured spectrum integral equal
to a thermal spectrum integrated over the same frequency band.  The
lowest temperature reached is $6.9 \pm 1.4$~mK.  Thus the cooling
factor from the ambient $T_M = 295$~K is $43000 \pm 11000$.
Systematic error in the calibration dominates statistical error in
these uncertainty estimates.  We note that the mechanical quality
factor was increased by a factor of about 80, from 19950 to
$1.6\times10^6$, by optical dilution.  Without an optical spring,
effective temperatures below 15~mK could not have been reached given
the mechanical losses of the oscillator.

\noindent {\bf Lifetime:} In this experiment, we began with
$\Omega_M = 2\,\pi \times 12.7$ Hz, $\Omega_M/\Gamma_M = 19950$, and
$T_M = 295$~K, while the coldest optical spring mode has
$\Omega_{\rm eff} = 2\,\pi \times 1018$~Hz, $\Omega_{\rm eff} /
\Gamma_{\rm eff} = 1.1$, and $T_{\rm eff} = 6.9 \times 10^{-3}$~K.
This corresponds to a factor of $196 \pm 40$ increase in $n_{\rm
osc}$; the error on this value comes from error estimates for
measured values of temperature, frequency and damping of the mode,
with the temperature uncertainty dominating.

Laser frequency noise presently limits the achievable degree of
cooling, but this noise source can be mitigated by placing two
identical cavities into the arms of a Michelson
interferometer~\cite{corbittPRA2006}. The laser light reflected from
each cavity interferes destructively at the beamsplitter, allowing
for rejection of laser noise at the antisymmetric output. The laser
frequency may be further locked to the common motion of the two
arms, providing additional stabilization of laser frequency noise.
The remaining differential motion of the arm cavity mirrors becomes
the oscillator degree of freedom to be placed in a quantum state.

The ultimate limit of optical cooling is expected to come from
vacuum noise of the optical field. The parameter regime required to
achieve cooling to the ground state with non-mechanical damping
techniques has been theoretically explored~\cite{CourtyEuroPhys2001,
vitaliPRA2002, vitaliJOSA2003,
  marquardtCondMat2007}, although the regime in which a low occupation
number may be reached with the aid of optical dilution has yet to be
delineated.  However, this will be the subject of another paper.

In conclusion, we have proposed a scheme that uses the optical
spring effect to both reduce the occupation number {\it and}
increase the dynamical lifetime of the mode of a 1 gram mirror
oscillator. We also provide an experimental demonstration showing
that cooling factors that exceed the mechanical $Q$ of the
macroscopic oscillator can be achieved when damping dilution from
the optical spring effect is used in conjunction with cold damping.

We would like to thank our colleagues at the LIGO Laboratory,
especially Rolf Bork and Jay Heefner, and the MQM group for
invaluable discussions. We gratefully acknowledge support from
National Science Foundation grants PHY-0107417 and PHY-0457264.

\end{document}